\documentclass{article}

\usepackage[english]{babel}

\usepackage[letterpaper,top=1cm,bottom=2cm,left=3cm,right=3cm,marginparwidth=1.75cm]{geometry}

\usepackage{amssymb}
\usepackage{biblatex}
\usepackage[ruled,vlined]{algorithm2e}
\usepackage{mathtools}
\usepackage{amsmath}
\usepackage{graphicx}
\usepackage[colorlinks=true, allcolors=blue]{hyperref}
\usepackage{xcolor}
\usepackage{authblk}

\title{On Energy-Dependent Neutron Diffusion}

\author[1,2]{Gabriele Burgio}
\author[1,2]{Christian Reiter}   
\author[3]{Stefano Lorenzi}

\affil[1]{TUM Chair of Applied Nuclear Technologies, Boltzmannstraße 15, Garching, Germany}
\affil[2]{Forschungs-Neutronenquelle Heinz Maier-Leibnitz (FRM II), Lichtenbergstraße 1, Garching bei M\"unchen, Germany}            
\affil[3]{Politecnico di Milano - Department of Energy, via La Masa 34 - 20156, Milano, Italy}

\begin{document}

\maketitle

\begin{abstract}

While the energy-dependent neutron diffusion equation is widely employed in nuclear engineering, its status as an approximation to the transport equation is not yet completely understood, and several different approximations are in use to determine the diffusion coefficients. Past work on the theory underlying the diffusion approximation has often made use of asymptotic arguments; in the energy-dependent case, however, papers have appeared that differ substantially in their findings. 
Here we present a formal asymptotic derivation of the multigroup diffusion equation which addresses these differences, along with the varying and sometimes physically stringent assumptions employed in these works. \newline 
Further, we show a way to exactly invert the relationship between flux and current in the $P_1$ approximation, giving a matricial expression for the multigroup diffusion coefficient which is formally exact, has clear physical meaning, and which can be easily computed to arbitrary precision on the basis of cross-section data already produced by lattice calculations. The resulting 2-group diffusion coefficient for an infinite medium of hydrogen is calculated with Monte Carlo, and compared to the those deriving from the Cumulative Migration Method and from the  out-scatter approximation. 

\end{abstract}





\section{Introduction}\label{sec:1}

Historically, beyond the simple $P_1$ approximation, diffusion and related $SP_N$ theories have been chiefly derived on the basis of either asymptotic or variational arguments \cite{Mclarren}. Fourier transform approaches have also been tried for very idealized formulations (see the introduction to \cite{Habetler}). \newline
Asymptotic arguments for diffusion have been introduced to neutron transport theory in the 70s by many authors (see \cite{larsen1980} and references therein). Several papers tackling the energy-dependent case have appeared, chiefly from Larsen and collaborators \cite{larsen_keller} \cite{larsen77} \cite{Morel} \cite{larsen}. \newline

However, the findings in these papers are sometimes incoungrent, both with each other and with the established computational practice. \newline
More specifically, many of these works \cite{larsen_keller} \cite{larsen77} \cite{larsen} obtain a diffusion equation for the leading order asymptotic solution by imposing solvability conditions on the second order equation. The resulting leading order solution is isotropic and separable in time and energy, $\psi_{0}(x,E,t) = A_0(x, t) \phi(x,E)$, and the diffusion equation is imposed only on the space-dependent part, with adjoint-weighed coefficients different from the ones used in practice. Underlying this separability is the physically stringent requirement that the medium under consideration have $k_{inf} = 1$. \newline
Other papers \cite{Mclarren} \cite{Morel} \cite{larsen} argue instead that an equation corresponding to commonly used multigroup diffusion determines a scalar flux which agrees with the integral of the exact angular flux up to second-order. They also make physically stringent assumptions, such as vanishing absorption and either infinite-medium criticality or negligible out-of-group scattering, and they don't provide a computable diffusion coefficient in the energy-dependent case. \newline

This work is composed of two parts: first, we present a simple formal derivation up to second order, whose purpose is to contextualize previous efforts and to address their physical meaning. Second, we derive an expression for the multigroup diffusion coefficient by exactly inverting the $P_1$ relation between flux and current. We argue that this shows that the strong assumptions that have been thus far employed in deriving multigroup diffusion asymptotically can be relaxed.  \newline
We believe that this latter part is also of practical value: the expression hasn't been previously published, to the authors' knowledge, and it provides a simple way to compute accurate multigroup diffusion coefficients to arbitrary precision, without the need for the many traditional approximations or specialized treatments \cite{CMM}. It also has a clear physical interpretation as an extension of the usual one-speed expression, accounting for the energy change due to scattering in the diffusion process. To conclude the paper, a simple numerical test is carried out for slowing down in hydrogen.
 
\section{Asymptotic Derivation of Time-dependent Diffusion}\label{sec:der}

\subsection{The Physical Scaling}

The traditional time-dependent diffusion equation corresponds to the asymptotic solution of a perturbation form of the neutron transport equation, which is singular with respect to both space and time \cite{Habetler}. Consequently, the flux would present initial and boundary layers, which determine the respective conditions. We do not analyze them here, rather choosing to focus on bulk phenomena. \newline
We consider the multigroup neutron transport equations in slab geometry:
\begin{equation}\label{eq:slab}
\begin{split}
    & \frac{\partial_{t} \psi_g(x, \mu)}{v_g} +  \mu \partial_{x} \psi_g(x, \mu) + \sigma_{t,g}(x)  \psi_g(x, \mu) =
    \\ & = \sum_{g'}  \int d\mu' {\sigma}_{s,g' \xrightarrow[]{} g}(x, \mu_0) \psi_{g'}(x, \mu')  + \frac{1}{2}  \sum_{g'} {\sigma}_{f,g' \xrightarrow[]{} g} \int d\mu' \psi_{g'}(x, \mu'). 
\end{split}
\end{equation}

We do not include delayed neutrons for simplicity, as they do not change the derivation qualitatively since their emission can be taken isotropic. \newline
To make the physical scalings of diffusion explicit, we multiply both sides of the system by the diagonal matrix $\Lambda_{g,g'} = \sigma_{t,g}^{-1} \delta^{g,g'}$, taking a characteristic value for each group for $\sigma_{t,g}$, to obtain:
\begin{equation}\label{eq:base_diff}
\begin{split}
   &  T_g \partial_{{t}} \psi_g(x, \mu) + \epsilon \mu \partial_{\hat{x}} \psi_g(x, \mu) + \hat{\sigma}_{t,g}(x)  \psi_g(x, \mu) =
   \\ & \sum_{g'}  \int d\mu' \hat{\sigma}_{s,g' \xrightarrow[]{} g}(x, \mu_0) \psi_{g'}(x, \mu')  + \frac{1}{2}  \sum_{g'} \hat{\sigma}_{f,g' \xrightarrow[]{} g} \int d\mu' \psi_{g'}(x, \mu') .
\end{split}
\end{equation}
Having defined:
\begin{equation}\label{eq:scaling}
\begin{split}
   &  \lambda_g / L = \epsilon;  ~~~ \lambda_g / v_g  = T_g; ~~~ \hat{x} = \frac{x}{L}
   \\ & \hat{\sigma}_{s,g' \xrightarrow[]{} g} :=  \sigma_{t,g}^{-1} {\sigma}_{s,g' \xrightarrow[]{} g}; ~~~ \hat{\sigma}_{f,g' \xrightarrow[]{} g} :=  \sigma_{t,g}^{-1} {\sigma}_{f,g' \xrightarrow[]{} g}
\end{split}
\end{equation}
where $L$ is a characteristic length for the flux, i.e. a length over which the fractional change in the flux (in any group and direction) is $O(1)$. $T_g$ is the mean free lifetime in group $g$. \newline

The scaling by $\epsilon$ of the spatial derivative term implies that the flux changes by a small fraction over a mean free path. With respect to a reactor, this corresponds to the established practice of using diffusion for homogenized cores, but not for shields. We note that it isn't necessary that absorption be low, in the system under consideration, for this condition to apply.
In general, the remaining factor in the spatial derivative term is now of the same order as the scaled reaction terms. \newline
We have not explicitly scaled the time derivative term to $\epsilon$ or $\epsilon^2$, as is customary. While the traditional scaling is both physically justified and mathematically necessary, in a strict sense, to obtain the diffusion equation rather than the telegrapher's equation, we mean to show that the infinite-medium criticality requirement that derives from it can be dispensed with under practical reactor conditions.  \newline
In the following we will imply the hat over the parameters, until the end of the section. \newline

\subsection{Asymptotic Solution}

We assume the following asymptotic expansion for the solution of equation (\ref{eq:base_diff}):
\[\psi_{g0}(x, \mu) = \psi_{g0}(x, \mu) + \epsilon \psi_{g1}(x, \mu) + \epsilon^2 \psi_{g2}(x, \mu) + o(\epsilon^2)\]

Consequently, at order zero it must hold:
\begin{equation}
    T_g \partial_t \psi_{g,0}(x, \mu) +  \sigma_{t,g}(x)  \psi_{g0}(x, \mu) = \sum_{g'}  \int d\mu' \sigma_{s,g' \xrightarrow[]{} g}(x, \mu_0) \psi_{g'0}(x, \mu')  + \frac{1}{2} \sum_{g'} \sigma_{f,g' \xrightarrow[]{} g} \int d\mu' \psi_{0g'}(x, \mu').
\end{equation}
That the solution $\psi_{g0}(x,t)$ is directionally homogeneous can be argued as follows: the equation above represents a situation in which there is no spatial coupling, so that, at any point, one effectively considers an infinite homogeneous system, where rotational invariance follows from the invariance of scattering with respect to incoming neutron direction. Integrating over angle then: 
\begin{equation}
   T_g \partial_t \psi_{g,00} +   \sigma_{t,g} \psi_{g,00} = \sum_{g'} \sigma_{s0, g' \xrightarrow[]{} g} \psi_{g',00} + \sigma_{f,g' \xrightarrow[]{} g} \psi_{g',00},   
\end{equation}
$\sigma_{s0, g' \xrightarrow[]{} g}$ and $\psi_{g,00}$ being zeroth Legendre moments for the respective quantities. \newline
We can see that, if we had scaled the time derivative by $\epsilon$ or $\epsilon^2$, we would have obtained the separability of the zeroth order solution: $\psi_{g,0}(x,t) = \psi_{g0, inf}(x) f(x,t) $, where the spectral part would be given by:
\begin{equation}\label{P0_homo}
   \sigma_{t,g}(x) \psi_{g0, inf} = \sum_{g'} \sigma_{s0, g' \xrightarrow[]{} g}(x) \psi_{g'0, inf} + \sigma_{f,g' \xrightarrow[]{} g}(x) \psi_{g'0, inf},   
\end{equation}
 Of course, this would imply that $k_{inf} = 1$. \newline
Further, considering a constant cross-section medium, (one can think of a homogenized assembly), $\psi_{g0, inf}$ wouldn't actually depend on $x$. This is physically reasonable, since a few mean free paths inside a homogeneous zone in a reactor a spectrum characteristic to the material will indeed be established \cite{Henry}, a fact which underlies the use of material buckling in spectrum calculations. \newline

At orders one and two we no longer have isotropy, on account of the streaming term, so we expand the solution in Legendre polynomials. The (trivial) first order condition on the $P_0$ component is:
\[\ \partial_{t} \psi_{g10} + \sigma_{t,g} \psi_{g10}   =  \sum_{g'} \left[ \sigma_{s0, g' \xrightarrow[]{} g} \psi_{g'10} +\sigma_{f,g' \xrightarrow[]{} g} \psi_{g'10} \right] .\]
The order one $P_1$ condition is:
\begin{equation}
     T_g \partial_t \psi_{g11} +  \sigma_{t,g} \psi_{g11} - \sum_{g'} \sigma_{s1, g' \xrightarrow[]{} g} \psi_{g'11} = - \frac{1}{3} \partial_x  \psi_{g00}.   
\end{equation}

At this point, we introduce an approximation in the treatment by neglecting the time derivative term in the $P_1$ condition for the following reason. \newline
It reasonable to assume, for a prompt-subcritical nuclear reactor, that the fractional change in current during a mean free lifetime $T_g < 10^{-4} s$ will be negligible when compared to the other terms, because both the cross-sections and the scale of spatial change have been normalized to $O(1)$ (cfr. \cite{Wigner} p. 233-234). Further, as will be seen, the time derivative of the current would only appear in the equation as a second order correction. We can thus justify the asymptotic derivation of multigroup diffusion for a medium where $k_{inf} \neq 1$, to a good approximation. \newline

We thus obtain an approximate $P_1$ condition:
\begin{equation}\label{eq:diff_def}
     \psi_{g11} \left[ \sigma_{t,g}  - \frac{\sum_{g'} \sigma_{s1, g' \xrightarrow[]{} g} \psi_{g'11}}{\psi_{g11}}  \right] = - \frac{1}{3} \partial_x  \psi_{g00}.   
\end{equation}

This relation, implicitly giving the expression for the diffusion coefficient, is found in the literature, both in derivations based on a $P_1$ expansion of the flux \cite{Glasstone} and in those employing asymptotic arguments (\cite{Morel} eq (50)). \newline
A central aspect of diffusion modeling is inverting this condition, in order to express the current in terms of the flux derivative as $\psi_{g11} = -D_g \partial_x  \psi_{g00} $, and effectively solve both $P1$ equations in one. \newline
Most often, the out-scatter approximation: 
\[ \sum_{g'} \sigma_{s1, g' \xrightarrow[]{} g} \psi_{g'11} \approx \sum_{g'} \sigma_{s1, g \xrightarrow[]{} g'} \psi_{g11}  \]
is employed, to obtain:
\begin{equation}\label{eq:out_scatter}
    \hat{D}_g = \frac{1}{3[\hat{\sigma}_{t,g} - \hat{\sigma}_{s0,g} \bar{\mu}_{0g}]}  \iff D_g = \frac{1}{3[\sigma_{t,g} - \sigma_{s0,g} \bar{\mu}_{0g}]}  .
\end{equation}
This approximation can be read as a detailed balance condition, which makes it valid in the limit of zero absorption \cite{Stammler}. It is also trivially valid in the limit of zero out-of-group scattering. \newline
We aim to show that these hypotheses aren't actually necessary for diffusion to physically take place, and we shall argue so in section (\ref{sec:diff_mean}) by inverting equation (\ref{eq:diff_def}) exactly, and discussing the physical meaning of the diffusion coefficient that is obtained. We will employ the out-scattering diffusion coefficient for now, as the new expression can be simply substituted for it.\newline
Other approaches to the treatment of (\ref{eq:diff_def}) are discussed in \cite{CMM}. \newline

At order two, using the recurrence relation for Legendre polynomials, the $P_0$ equation is:
\[ \partial_{t} \psi_{g20} + \partial_x \psi_{g11} + \sigma_{t,g} \psi_{g20}   =  \sum_{g'} \left[ \sigma_{s0, g' \xrightarrow[]{} g} \psi_{g'20} +\sigma_{f,g' \xrightarrow[]{} g} \psi_{g'20} \right] .\]

Finally, using the first order $P_1$ equation and summing the $P_0$ equation at all orders, multiplied by 1, $\epsilon$ and $\epsilon^2$ respectively, we obtain:
\begin{equation}
\begin{split}
   &   T_g( \partial_{t} \psi_{g,00} + \epsilon \partial_{t} \psi_{g,10} + \epsilon^2 \partial_{t} \psi_{g,20}) + \hat{\sigma}_{t,g} (\psi_{g,00} + \epsilon \psi_{g,10} + \epsilon^2  \psi_{g,20}  ) - \epsilon^2 \partial_{\hat{x}} \hat{D}_g \partial_{\hat{x}} \psi_{g, 00} =
   \\ & \sum_{g'} \hat{\sigma}_{s0,g' \xrightarrow[]{} g} ( \psi_{g',00} + \epsilon \psi_{g',10} + \epsilon^2 \psi_{g', 20}) +  \sum_{g'} \hat{\sigma}_{f,g' \xrightarrow[]{} g} (\psi_{g',00} + \epsilon \psi_{g',10} + \epsilon^2 \psi_{g', 20})
\end{split}
\end{equation}
where we have recovered the hats. \newline
Simply integrating the expansion for the angular flux, we can define the scalar flux $\Phi_g := ( \psi_{g,00} + \epsilon  \psi_{g,10} + \epsilon^2 \psi_{g,20})$, so that it holds:
\begin{equation}
\begin{split}
   &  T_g \partial_{t}\Phi_g  + \hat{\sigma}_{t,g} \Phi_g - \epsilon^2 \partial_{\hat{x}} \hat{D}_g \partial_{\hat{x}} \Phi_g =  \sum_{g'} \hat{\sigma}_{s0,g' \xrightarrow[]{} g} \Phi_{g'} +  \sum_{g'} \hat{\sigma}_{f,g' \xrightarrow[]{} g} \Phi_{g'} 
\end{split}
\end{equation}
With an error which is $o(\epsilon^2)$. \newline
Now we can multiply this equation by $\Lambda^{-1}$ and substitute back the scaling relationships and definitions given in (\ref{eq:scaling}). In particular, for the diffusion coefficient term, it holds:
\[\hat{D}_g = \sigma_{t,g} D_g; ~ ~  \partial_{\hat{x}} = L \partial_x ~  \Longrightarrow ~  \sigma_{g,t} \epsilon^2 \partial_{\hat{x}} \hat{D}_g \partial_{\hat{x}} = \partial_x D_g \partial_x   \]
so that we have recovered the multigroup diffusion equation. \newline

\section{The Multigroup Diffusion Coefficient}\label{sec:diff_mean}

\subsection{Exact Solution of the $P_1$ Current Condition}

The out-scatter approximation presented in equation (\ref{eq:out_scatter}) is not a good one when energy changes due to anisotropic scattering (in the laboratory system) are not negligible. This is a rather common occurrence in nuclear engineering, for example on account of scattering by hydrogen. Better approaches to the solution of equation (\ref{eq:diff_def}) have therefore been the object of much study in recent decades \cite{Henry} \cite{CMM}. \newline In this section, we aim to show that the relation (\ref{eq:diff_def}) can actually be inverted exactly, giving rise to an expression that is very amenable to numerical calculation. \newline
We can restate the diffusion coefficient equation (\ref{eq:diff_def}) as follows:
\begin{equation}\label{eq:base}
    \sum_{g'} \left[ \delta^{g,g'}\sigma_{t,g'}  - \sigma_{s1, g' \xrightarrow[]{} g} \right]\psi_{g'11} = - \frac{1}{3} \partial_x  \psi_{g00} .
\end{equation}
We now define the matrices $T_{g,g'} = \delta^{g,g'}\sigma_{t,g'}$ and $S_{g,g'} = \sigma_{s1, g' \xrightarrow[]{} g}$ to rewrite (\ref{eq:base}) in matrix form:
\[ \left[ T  - S \right]\psi_{11} = - \frac{1}{3} \partial_x  \psi_{00} . \]
It follows that:
\[ \psi_{11} =  (Id - T^{-1} S)^{-1} T^{-1} (- \frac{1}{3} \partial_x  \psi_{00}) = \left[ \sum_{n\geq0} (T^{-1} S)^n  \right] T^{-1} (-\frac{1}{3} \partial_x  \psi_{00})\] 
where the existence of the inverse and the second equality are given by the theorem of the Neumann series, which holds if the series $\sum_{n \geq 0} (T^{-1}S)^n $ converges \cite{Kato}.   We shall now show that this is the case, while also clarifying the physical meaning of the expression. \newline

To start, let us consider $\sigma_{s1, g' \xrightarrow[]{} g}$, which is the first Legendre moment of the scattering cross-section $\sigma_{s, g' \xrightarrow[]{} g}(\mu)$. One can always normalize such a function of $\mu$ over $[-1,1]$ as follows:
\[\sigma_{s, g' \xrightarrow[]{} g}(\mu) = \sigma_{s0, g' \xrightarrow[]{} g} W_{g' \xrightarrow[]{} g}(\mu)  \]
where $\sigma_{s0, g' \xrightarrow[]{} g}$ is the total $g'$-to-$g$ scattering cross-section (the total corresponds to the zeroth moment because cross-sections are nonnegative functions). \newline
It follows then, for the average scattering cosine from $g'$ to $g$, which we call $\mu_{0, g' \xrightarrow[]{} g}$, that:
\begin{equation}\label{eq:cosine}
   \mu_{0, g' \xrightarrow[]{} g} = \int_{-1}^{1} d\mu W_{g' \xrightarrow[]{} g}(\mu) \mu =  \frac{\sigma_{s1, g' \xrightarrow[]{} g} }{\sigma_{s0, g' \xrightarrow[]{} g} }  ~ \Longrightarrow ~  \sigma_{s1, g' \xrightarrow[]{} g} = \mu_{0, g' \xrightarrow[]{} g} \sigma_{s0, g' \xrightarrow[]{} g} .
\end{equation}

It holds, for the nth term in the Neumann series:
\[(T^{-1} S)^n = T^{-1} P^{n-1} S \]
where:
\[ P_{g,g'} :=  (S T^{-1})_{g,g'}  = \frac{1}{\sigma_{t,g'}} \sigma_{s1, g' \xrightarrow[]{} g} .\]
If there is nonzero absorption probability in each energy group, as is physically the case outside of vacuum, then clearly $\| P \| <1$ as an operator over vectors of size $G$ (there are fewer neutrons scattered from $g$ to other groups than are lost in $g$ overall).\newline
The same holds, for whatever absorption probability, unless the average scattering cosine is 1 (in which case scattering does not actually change neutron direction). \newline
Because $T^{-1}$ and $S$ are bounded, $\| (T^{-1} S)^n \|< C \| P \|^{n-1}$, and the series converges.  

\subsection{Physical Interpretation of the Diffusion Coefficient}

Now we want to showcase how the obtained diffusion coefficient represents the movement of neutrons resulting from the consecutive scatterings that cause the Brownian motion. \newline 
To do so, we write out the action of the first few terms of the series that defines the operator on a generic vector $y_g$:
\[ [T^{-1} y]_g  = \lambda_g y_g \]
\[ [(T^{-1} S) T^{-1} y]_g  =  \lambda_{g} \sum_{g'} \frac{\sigma_{s1, g'\xrightarrow[]{} g}}{\sigma_{t,g'}}  y_{g'} \]
\[ [(T^{-1} S)^2 T^{-1} y]_g  =  \lambda_{g} \sum_{g''} \frac{\sigma_{s1, g'' \xrightarrow[]{} g}}{\sigma_{t,g''}} \sum_{g'} \frac{\sigma_{s1, g' \xrightarrow[]{} g''}}{\sigma_{t,g'}}  y_{g'} \]
and so on. \newline
It is now easy to see that the diffusion coefficient that has been found describes a process of repeated scatterings, where the overall traveled distance equals the sum over the whole neutron lifetime of the mean free path at the current energy, weighed by the likelihood of having been scattered to the current energy rather than absorbed, and by the average change in direction (given by the cosine in eq. (\ref{eq:cosine})) that took place with the scattering to the current energy. Fission breaks the chain, as is reasonable since it is assumed isotropic.\newline
We can therefore define:
\begin{equation}\label{eq:lambda_def}
   \tilde{\Lambda}_{g,g'} := \left( \Lambda \sum_{n\geq0} P^n \right)_{g,g'}; ~~~ D_{g,g'} := \frac{\tilde{\Lambda}_{g,g'}}{3}   
\end{equation}
where $\tilde{\Lambda}_{g,g'}$ is the matrix containing the average total migration length in energy group $g$ throughout the lifetime of a neutron starting in energy group $g'$. \newline
The given expression for the energy-dependent diffusion coefficient can be computed to arbitrary precision through the series, which converges exponentially. The necessary order of $n$ to be considered for convergence is given by the number of scatterings before absorption in the system, but it isn't practically onerous to include many more. \newline 
Our derivation starts form the multigroup transport equation, where in general all cross-sections (including $\sigma_t$) should depend on the angle, on account of the implicit weighing. 
However, the steps relevant to the new diffusion coefficient could be carried out in the same way in continuous energy, simply substituting integrals to the sums (the operator norm points holds in the same way). Further, flux angular dependence is negligible inside large homogeneous regions, as are often of interest for diffusion calculations in any case. \newline
The limiting case for the scattering angle $\mu_{0, g' \xrightarrow[]{} g} = 1$ is generally taken to prevent diffusion, and it does so in our case as well (assuming vanishing absorption, as is done in \cite{Morel}), since in that case the $P$ matrix no longer has norm less than one. However, a diffusion coefficient can be defined even for this limiting case if there is non-negligible absorption, although it should be borne in mind that we are also requiring that the flux change slowly in space on the scale of a mean free path. Practically this could be the case if fission and absorption roughly balance each other out inside a reactor region. \newline

\subsection{Comparison with Other Diffusion Coefficients}

As a basic check for the obtained expression, a Serpent 2 \cite{Serpent} calculation was ran representing an isotropic source of Watt-spectrum neutrons inside an infinite homogeneous medium, composed entirely of hydrogen, using reflective boundary conditions. A total population of $10^7$ neutrons was used. \newline
The values of the multigroup diffusion coefficient were computed according to expression (\ref{eq:lambda_def}) in the 172 group structure of WIMS and then collapsed into a 2-group matrix, flux-weighing the transport migration length. This is compared with the Cumulative Migration Method \cite{CMM} and out-scattering diffusion coefficients, both collapsed from the same fine grid, in the following: 
\[
\begin{array}{ccc}
D_{CMM} =
\begin{pmatrix}
0.1330 & 0 \\
0 & 0.01108
\end{pmatrix}
&
D_{OS} =
\begin{pmatrix}
0.1914 & 0 \\
0 & 0.01077
\end{pmatrix}
&
D_{MG} =
\begin{pmatrix}
0.1898 &  3e-08\\ 
0.00088 & 0.01076
\end{pmatrix}
\end{array}
\]
The CMM has been shown to perform well for the thermal group diffusion coefficient \cite{CMM}, therefore it can be used as a reference for thermal neutron migration. It can be seen that our expression predicts a lower thermal-to-thermal value, but the fast-to-thermal coefficient corrects this. Because the fast flux fraction is about $0.3$ in this benchmark, it appears that the predicted thermal leakage will be nearly equal. More thorough testing is foreseen, which will however require an examination of either boundary or interface conditions. 

\section{Conclusions}
In this article, an asymptotic derivation for the multigroup neutron diffusion equation has been presented. It was shown that vanishing absorption and low out-of-group scattering are not necessary conditions for said equation to hold, at least as long as the relation giving the transport cross-section is inverted exactly. The expression that was derived for the diffusion coefficient can be simply calculated in both Monte Carlo and deterministic lattice codes and might prove a more straightforward alternative to the in-scatter method after more thorough testing. It seems especially well-suited for Monte Carlo cross-section generation, as the strength of the method in treating spatial heterogeneity can be harnessed directly. \newline

For the future, more thorough benchmarking of the multigroup diffusion coefficient on complex heterogeneous models is foreseen. Likewise, a more complete assessment of the effect and treatment of the angular dependence of the total cross-section in the multigroup case is desirable. \newline
Extensions of interest for the theoretical part would be to the treatment of higher-order contributions to the initial and boundary layers in the energy-dependent case. 

\section*{Acknowledgements}

This work was supported by a combined grant (FRM2427) from the Bundesministerium für Forschung, Technologie und Raumfahrt (BMFTR) and the Bayerisches Staatsministerium fur Wissenschaft und Kunst (StMWK). \newline
The authors want to thank Professor Edward Larsen for his remarkable availability and patience in discussing technical matters.

\printbibliography

\end{document}